\documentclass[aps,pre,twocolumn,groupedaddress,showpacs]{revtex4}
\usepackage{graphicx}
\begin{document}

\title{Strain Hardening of Polymer Glasses: Entanglements, Energetics, and Plasticity}
\author{Robert S. Hoy}
\email{robhoy@mrl.ucsb.edu}
\author{Mark O. Robbins}
\affiliation{Department of Physics and Astronomy, Johns Hopkins University, Baltimore, MD 21218}

\date{January 5, 2008}
\begin{abstract}
Simulations are used to examine the microscopic origins of strain
hardening in polymer glasses.
While stress-strain curves for a wide range of temperature can be fit
to the functional form predicted by entropic network models,
many other results
are fundamentally inconsistent with the physical
picture underlying these models.
Stresses are too large to be entropic and have the wrong trend
with temperature.
The most dramatic hardening at large strains reflects increases
in energy as chains are pulled taut between entanglements rather
than a change in entropy.
A weak entropic stress is only observed in shape recovery of deformed
samples when heated above the glass transition.
While short chains do not form an entangled network,
they exhibit partial shape recovery, orientation, and strain hardening.
Stresses for all chain lengths collapse when plotted against
a microscopic measure of chain stretching rather than the macroscopic
stretch.
The thermal contribution to the stress is directly proportional to
the rate of plasticity as measured by breaking and reforming of
interchain bonds.
These observations suggest that the correct
microscopic theory of strain hardening should be based on glassy state
physics rather than rubber elasticity.
\end{abstract}
\maketitle

\section{Introduction}
\label{sec:intro}

Glass forming polymers are of great industrial importance and scientific
interest because of their unique mechanical properties, which arise from the
connectivity and random-walk-like structure of the constituent chains.
At large strains, the stress increases as the chain molecules orient,
in a process known as strain hardening.
Strain hardening suppresses strain localization (crazing, necking,
shear banding) and is critical in determining material properties such as
toughness and wear resistance.

Traditional theories of glassy strain hardening \cite{haward68,arruda93b}
assume that polymer glasses behave like crosslinked rubber, with the number
of monomers between crosslinks equal to the entanglement length $N_e$.
The increase in the stress $\sigma$ due to deformation by a stretch
tensor $\bar \lambda$ is attributed to the decrease in entropy as
polymers stretch affinely between entanglements.
Beyond the initial plastic flow regime, entropic
network models predict \cite{arruda93b}:
\begin{equation}
\sigma(\bar\lambda) = \tau_{flow} + G_{R}g(\bar\lambda)L^{-1}(h)/3h
\label{eq:langevinmodel}
\end{equation}
where $\tau_{flow}$ is the plastic flow stress, $G_{R}$ is the strain
hardening modulus, $L^{-1}$ is the inverse Langevin function,
$g(\bar\lambda)$ describes the entropy reduction for ideal
Gaussian chains, and $L^{-1}(h)/3h$ corrects for the finite length
of segments between entanglements.
The value of $N_{e}$ enters in $h$, which is the ratio
of the Euclidean distance between entanglements to the contour length.

Entropic network models have had much success in fitting experimental
data \cite{arruda95,arruda93b,boyce94,ping99}.
However, serious discrepancies between the models and experiments are
revealed in trends with temperature and the values of fit parameters
\cite{kramer05}.
Entropic models predict $G_R = \rho_e k_{B} T$, where $\rho_{e}$ is the
entanglement density.
This is about 100 times smaller than values measured near
the glass transition temperature $T_{g}$
\cite{vanMelick03}.
Moreover,
$G_R$ increases monotonically as $T$ decreases \cite{vanMelick03,hoy06},
while any entropic stress must drop to zero as $T \rightarrow 0$.
Parameters such as $G_R$ and $N_e$, which entropic models assume to be
material constants, must be adjusted significantly to fit data for
different strain states (i.e. uniaxial or plane strain) \cite{gsell94}.
Fits to
the shape of strain hardening curves also yield smaller values of $N_e$
than those
obtained from the plateau moduli of melts \cite{arruda93b}.

A more fundamental conceptual difficulty with entropic models is that,
unlike rubber, glasses are not ergodic.
For $T < T_g$, thermal activation is not sufficient to allow chains to
sample conformations freely.
Rearrangements occur mainly under active deformation \cite{capaldi02,loo00}
and at a frequency that scales with the strain rate \cite{chen07}.
In such far from equilibrium situations, the relevance of
entropy is unclear.
In addition,
experiments \cite{hasan93,rittel99,li01}
and simulations \cite{chui99,hoy07,li06}
show that the internal energy contributes to strain hardening,
but this is not included in entropic network models.

Molecular simulations allow a full analysis of the mechanisms
of large strain deformation in glassy polymers
\cite{brown91,mckechnie93,mott93,jang99,capaldi02,capaldi04,fortunelli04,lyulin04,lyulin05,li06}.
In recent papers \cite{hoy06,hoy07},
we have examined the origins of strain hardening
using a coarse-grained bead-spring model \cite{kremer90}.
As in experiments, numerical values of $G_R$ are much 
larger than predicted by entropic models and show the opposite
trend with temperature \cite{hoy06}.
A direct correlation between $\tau_{flow}$ and $G_R$ was
discovered that allowed curves for different interactions,
strain rates and temperatures to be collapsed \cite{hoy06}.
Substantial strain hardening was observed for chains that
are shorter than $N_e$ and thus do not form a network \cite{lyulin04,hoy07}.
For chains of all lengths, strain hardening was directly
related to strain-induced orientation and the rate of
plastic rearrangement \cite{hoy07}.

This paper extends our simulation studies in several directions.
Uniaxial and plane strain compression are examined
for a wide range of $N_e$, $T$ and chain lengths.
Stress curves for all entangled systems 
can be fit to Eq. \ref{eq:langevinmodel}.
The fits confirm the connection between $\tau_{flow}$ and $G_R$ \cite{hoy06},
which both drop linearly to zero as $T$ rises to $T_g$.
This observation motivates a modification of Eq. \ref{eq:langevinmodel}.
Using $T_g$ and the fit to a stress-strain curve at one
temperature, the model predicts strain hardening curves
for all $T<T_g$ remarkably well.
However, as in experiments \cite{gsell94},
it is necessary to vary parameters such as $G_{R}$ and $N_e$ in
unphysical ways in order to
fit curves for different strain states.

Direct examination of entropic and energetic contributions to strain
hardening reveals qualitative failures of network models.
The rapid hardening at large strains that is fit by the Langevin
correction in Eq. \ref{eq:langevinmodel}
does not reflect a reduction in entropy.
Instead there is a rapid rise in energetic stress
as chains are pulled taut between entanglements.
Variation in the energetic contribution for different
strain states leads to changes in fit values of $N_e$.
For all chain lengths and $N_e$, we find that the thermal part of the
stress correlates directly with breaking and reformation of van der Waals
bonds during deformation.
This provides an explanation for the correlation between $G_R$ and
$\tau_{flow}$.

Remarkable shape recovery is observed in experiments when highly
deformed samples are unloaded and heated slightly above $T_{g}$ \cite{haward97}.
Network models assume this recovery is driven by a ``back stress''
related to the entropy of the entanglement network, and shape recovery
is often seen as providing evidence for entropic strain hardening.
Our simulations also show dramatic shape recovery that is driven
by orientational entropy.
However, the magnitude of the associated stress is only of order $\rho_e k_B T$
and thus much too small to be significant in strain hardening.

Changes in microscopic chain conformations are also explored.
While Eq. \ref{eq:langevinmodel} assumes affine deformation of segments
of length $N_e$, the observed deformation
becomes increasingly subaffine as strain increases.
This reflects a straightening of segments between entanglements that
disturbs the local structure of the glass and increases the internal energy.
Although unentangled chains do not form a network, they still
undergo significant orientation during strain \cite{lyulin04,hoy06}.
For all chain lengths, the thermal contribution to
the stress is directly related to the orientation
of chains on the end-to-end scale rather than the macroscopic
stretch \cite{ken,hoy07}.

The following section describes the potentials, geometry and strain
protocol used in our simulations.
Next, fits to entropic network models are examined, and an extension
that incorporates the correlation between $\tau_{flow}$ and
$G_R$ is presented.
This is followed by a discussion of the energetic and entropic
contributions to the stress and the role of entropic back
stresses in shape recovery.
Sections \ref{subsec:chainlengthdep} and \ref{subsec:dissiplastic} explore the effect
of chain length and orientation and demonstrate the connection between
plastic deformation and the thermal component of the stress.
The final section presents a summary and conclusions.

\section{Polymer Model and Methods}
\label{sec:modelmethod}

We employ a coarse-grained bead-spring polymer model \cite{kremer90} that incorporates key physical features of linear homopolymers such as covalent backbone bonds, excluded-volume and adhesive interactions, chain stiffness, and the topological restriction that chains may not cross.
Each linear chain contains $N$ spherical monomers of mass $m$.
All monomers interact via the truncated and shifted 
Lennard-Jones potential:
\begin{equation}
U_{LJ}(r)  =  4u_{0}\left[ \left(\frac{a}{r}\right)^{12}-
\left(\frac{a}{r}\right)^{6}-
\left(\frac{a}{r_{c}}\right)^{12}+
\left(\frac{a}{r_{c}}\right)^{6}\right],
\label{eq:shiftedLennardJones}
\end{equation}
where $r_{c}$ is the potential cutoff radius and
$U_{LJ}(r) = 0$ for $r > r_{c}$.
We express all quantities in terms of the molecular diameter $a$, energy scale $u_{0}$, and characteristic time $\tau_{LJ} = \sqrt{ma^{2}/u_{0}}$. 

Covalent bonds between adjacent monomers on a chain are modeled using the finitely extensible nonlinear elastic (FENE) potential 
\begin{equation}
\begin{array}{lll}
U_{FENE}(r) & = & -\displaystyle\frac{kR_{0}^2}{2}\rm ln(1 - (r/R_{0})^{2})\ \ ,
\end{array}
\label{eq:FENEbondpotential}
\end{equation}
with the canonical parameter choices \cite{kremer90} $R_{0} = 1.5a$ and $k = 30u_{0}/a^{2}$.
The equilibrium bond length $l_{0} \simeq 0.96a$.
The FENE potential does not allow chain scission, but the maximum
tensions on covalent bonds for the systems studied here are well below the
critical value for scission in breakable-bond models \cite{rottler03}. 

As a means of varying $N_{e}$, we introduce chain stiffness using the bending potential
\begin{equation}
U_{bend}(\theta) = k_{bend}(1 - cos\theta)\ \ ,
\label{eq:ChainStiffnessPotential}
\end{equation}
where $\theta$ is the angle betweeen consecutive covalent bond vectors
along a chain.
Increasing $k_{bend}$ increases the root-mean-squared (rms)
end-to-end distance of chains $R_{ee}$.
The chain stiffness constant $C_{\infty} \equiv <R_{ee}^2>/(N-1)l_0^2$ for
well-equilibrated \cite{auhl03} melt states rises from $1.8$ to $3.34$
as $k_{bend}$ is increased from $0$ to $2.0u_{0}$.
The value of $N_{e}$ decreases from about 70 to about 20 over the same interval \cite{everaers04}.  
The key parameter in entropic network models is the number of statistical
segments per entanglement $N_e/C_{\infty}$ (Table \ref{tab:systemparameters}).

The initial simulation cell is a cube of side length $L_{0}$, where $L_{0}$ is greater than the typical end-to-end distance of the chains. 
$N_{ch}$ chains are placed in the cell, with periodic boundary conditions applied in all three directions. 
Each initial chain configuration is a random walk of $N-1$ steps with the distribution of bond angles chosen to give the targeted large-scale equilibrium chain structure.
In particular the mean value of $cos(\theta)$ is adjusted so that 
\begin{equation}
C_{\infty} = \displaystyle\frac{1 + <cos(\theta)>}{1 - <cos(\theta)>} \ \ .
\label{eq:chainstiffnessconst}
\end{equation}
$N_{ch}$ is chosen so that the total number of monomers $N_{tot}=NN_{ch}$ is $250000$ $(L_0 = 66.5a)$ for flexible ($k_{bend} = 0$) chains  and $70000$ $(L_0 = 43.5a)$ for semiflexible ($k_{bend} > 0$) chains.
The initial monomer number density is $\rho = 0.85a^{-3}$.

After the chains are placed in the cell, we perform molecular dynamics (MD) simulations.
Newton's equations of motion are integrated with the velocity-Verlet method 
\cite{frenkel02} and timestep $\delta t = .007\tau_{LJ}-.012\tau_{LJ}$.  
The system is coupled to a heat bath at temperature $T$ using a Langevin 
thermostat \cite{schneider78} with damping rate $1.0/\tau_{LJ}$.
Only the peculiar velocities are damped.

We first equilibrate the systems thoroughly at $T = 1.0u_{0}/k_{B}$,
which is well above the
glass transition temperature $T_{g} \simeq 0.35u_{0}/k_{B}$ \cite{rottler03c}. 
The cutoff radius $r_{c}$ is set to $2^{1/6}a$, as is standard in 
simulations of melts with the bead-spring model \cite{kremer90,everaers04}.
For well-entangled chains, the time required for diffusive equilibration is prohibitively large.
To speed equilibration we use the double-bridging-MD hybrid (DBH) algorithm \cite{auhl03}, where Monte Carlo moves that alter the connectivity of chain subsections are periodically performed. 

Glassy states are obtained from well-equilibrated melts by performing a rapid temperature quench at a cooling rate of  
$\dot{T} = -2\times10^{-3}u_{0}/k_{B}\tau_{LJ}$.
We increase $r_{c}$ to its final value, typically $1.5a$, and cool at constant
density until the pressure is zero.
The quench is then continued at zero pressure using a Nose-Hoover barostat \cite{frenkel02} with time constant 10$\tau_{LJ}$ until the desired $T$ is reached.
Larger values of $r_{c}$ lead to higher final densities and larger stresses at all strains \cite{rottler01}, but we have checked that using values of $r_{c}$ as large as $2.6a$ does not change the conclusions presented below. 
Indeed, stress-strain curves for different $r_c \geq 1.5a$ collapse when normalized by $\tau_{flow}$ \cite{hoy06}.
In this paper $T$ varies from 0 to $0.3u_0/k_B$.
Simulations at $T=0$ are not directly relevant to experiments, but are useful to gain theoretical understanding of polymer deformation in the limit where thermal activation is not important.
To operate in the $T\to0$ limit, we remove the Gaussian noise term from the standard Langevin thermostat and retain the viscous drag term. 

Values of $N_{e}$ (Table \ref{tab:systemparameters}) are measured by performing primitive path analyses (PPA) \cite{everaers04,zhou05}.
Details of the PPA procedure are the same as those used for undiluted systems in our recent paper \cite{hoy06}.  
Melt entanglement lengths are consistent with values of $N_{e}$ from the melt plateau moduli \cite{everaers04}. 
Quenching melt states into a glass has little effect on the values of $N_{e}$ determined from PPA \cite{hoy05}.
The changes in entanglement density $\rho_{e} = \rho/2N_{e}$ upon cooling are primarily due to a 15\% increase in $\rho$.
The conclusion that glasses inherit the melt value of $N_{e}$ is consistent with experimental \cite{kramer83} and simulation \cite{rottler03} studies of the craze extension ratio, as discussed in Section \ref{subsec:modeightchain}.
However, it is inconsistent with some entropic models of strain hardening  \cite{arruda95} that assume that $\rho_{e}$ increases rapidly as $T$ decreases.

The values of $N$ employed in this paper are $12-500$ for flexible chains and $4-350$ for semiflexible chains,  spanning the range from
the unentangled to the fully entangled $(N \gg N_{e})$ limits.
It is important to note that unentangled systems $(N<N_e)$ are
often brittle.
This may severely limit the maximum strain that can be studied
in experiments and complicate comparison to our simulations.

\begin{table}[h]
\caption{Chain statistics in fully entangled glasses $(N/N_e > 7)$}
\begin{ruledtabular}
\begin{tabular}{lccc}
$k_{bend}a^2/u_0$ & $N_e$ & $C_{\infty}$ & $N_e/C_{\infty}$\\
0.0 & 71 & 1.70 & 42\\
0.75 & 39 & 2.05 & 19\\
1.5 & 26 & 2.87 & 9\\
2.0 & 22 & 3.29 & 7\\
\end{tabular}
\end{ruledtabular}
\label{tab:systemparameters}
\end{table}

In fundamental studies of strain hardening \cite{vanMelick03, arruda93b,haward93}, compressive rather than tensile deformation is preferred because it suppresses strain localization.
This allows the stress to be measured in uniformly strained systems.
The rapidity of the quench used here minimizes strain softening, which in turn yields ductile, homogeneous deformation even at the lowest temperatures and highest strains considered. 

We employ two forms of compression; uniaxial and plane strain.
The stretch $\lambda_{i}$ along direction $i$ is defined as $L_{i}/L_{i}^{0}$, where $L_i^{0}$ is the cell side length at the end of the quench.
In uniaxial compression, the systems are compressed along one direction, $z$, while maintaining zero stresses along the transverse $(x,y)$ directions \cite{yang97}. 
In plane strain compression, the systems are also compressed along the $z$ direction, and zero stress is maintained along the $x$ direction, but the length of the system along the $y$ direction is held fixed ($\lambda_{y} = 1$).

Compression is performed at constant true strain rate $\dot{\epsilon} = \dot{\lambda}_{z}/\lambda_{z}$, which is the favored protocol for strain hardening experiments \cite{haward93}.
The systems are compressed to true strains of $\epsilon = -1.5$,
corresponding to  $\lambda_{z} = \exp(-1.5) \simeq 0.22$, for uniaxial compression and $\epsilon = -1.2$ ($\lambda_z \simeq 0.30$)
for plane strain compression.
These strains are similar to the highest achievable experimentally \cite{hasan93} in glassy state compression. 

Simulations were performed
at $\dot{\epsilon}$ between $-10^{-5}/\tau_{LJ}$ and $-10^{-3}/\tau_{LJ}$.
As in previous studies of strain hardening \cite{hoy06} and the initial
flow stress \cite{rottler03c,rottler05}, 
a weak logarithmic rise in stress with strain rate was observed
for $|\dot{\epsilon}| \lesssim 3 \cdot 10^{-4}$.
This small rise does not change the conclusions drawn in the following
sections and a similar logarithmic rise is observed in many experiments
\cite{mulliken06}.
Thus, while our simulations are performed at much higher strain rates
than experiments, we expect that they capture experimental trends.
A more rapid change in behavior was observed for
$|\dot{\epsilon}| \gtrsim 3 \cdot 10^{-4}$
and qualitative changes in behavior can occur at the much 
higher rates employed in some previous simulations 
\cite{lyulin04,lyulin05}.
For example, the time for stress equilibration across the sample
is of order $L_0/c_s$ where $c_s$ is the lowest sound velocity.
When this time is larger than the time between plastic rearrangements,
then each rearrangement occurs before the stress field around it has
fully equilibrated in response to surrounding rearrangements.
The decorrelated relaxation of different regions leads to a more
rapid rise in yield stress with rate \cite{rottler03c}.
The effect of rate on the relaxation of unentangled chains is discussed
in Section \ref{subsec:chainlengthdep}.

\section{Results}
\label{sec:results}

\subsection{Comparison to Eight-Chain Model}
\label{subsec:modeightchain}

The eight chain model \cite{arruda93b} has been very successful in describing the functional form of stress-strain curves and is widely used to fit experimental results \cite{haward97}. 
It assumes that the entanglement network deforms affinely at constant volume and employs a body centered cubic network geometry with eight chains per node.
The stretch of each segment between nodes is then $\lambda_{chain} = ((\lambda_{x}^{2} + \lambda_{y}^{2} + \lambda_{z}^{2})/3)^{1/2}$, yielding $h = \lambda_{chain}/\sqrt{N_{e}/C_{\infty}}$ in Eq.\ \ref{eq:langevinmodel}.
This choice of network was the main advance of the eight chain model over previous entropic models \cite{haward68, boyce88}. 
It allows the model to fit stress-strain curves for various strain states, i.e.\ shear \cite{boyce94, ping99} and uniaxial \cite{arruda93b, arruda95} or plane strain  \cite{arruda93b,boyce94} compression
\cite{foot1}.
The prediction for the difference between principal stresses along axes $i$ and $j$ in the strain hardening regime is
\begin{equation}
\sigma_{i} - \sigma_{j} = \tau^{ij}_{flow} + G_R \frac{L^{-1}(h)}{3h}
( \lambda_{i}^{2} - \lambda_{j}^{2} ) \ \ ,
\label{eq:origeightchain}
\end{equation}
where $\tau^{ij}_{flow}$ is an independently modeled,  rate- and temperature-dependent plastic flow stress \cite{boyce88, mulliken06}.

Equation \ref{eq:origeightchain} simplifies for the cases of uniaxial and
plane strain compression considered here and in many experiments.
For uniaxial strain only $\sigma_z$ is nonzero and the constant volume
constraint implies $\lambda_x=\lambda_y=\lambda_z^{-0.5}$.
For plane strain compression, the constant volume constraint requires
$\lambda_x=\lambda_z^{-1}$ and both $\sigma_z$ and $\sigma_y$ are
nonzero.
Equation \ref{eq:origeightchain} implies a relation between the strain
hardening of $\sigma_{z}$ and $\sigma_{y}$, but the latter does not appear
to have been measured in experiments.

Despite its wide use, there are fundamental difficulties with the eight chain model that were noted in the Introduction.
As a rubber-elasticity based model, it predicts $G_{R} = \rho_{e}k_{B}T$.
This prediction is about 100 times too small at $T\sim0.9T_{g}$
if values of $\rho_{e}$ are estimated from the melt plateau modulus
and has the wrong trend with decreasing $T$ \cite{vanMelick03,kramer05}.
Some models \cite{arruda95,richeton07} assume $\rho_e$ is much larger than in the melt
and rises rapidly as $T$ decreases below $T_g$ in order to fit experiments.
However, studies of crazing in polymer glasses
do not indicate any increase in $\rho_e$ over the melt \cite{kramer83,rottler05}.
A constant entanglement density is also consistent with the idea that
entanglements represent topological constraints and the observation
that the topology does not evolve significantly below $T_g$.
The extra entanglements added in network models as $T$ decreases
may capture the effect of glassy constraints associated with energy barriers,
but it is not clear that it is natural to treat such constraints
within a rubber-elasticity framework.

Another shortcoming of the eight chain model and more recent work \cite{richeton07,mulliken06} is that the flow stress must be introduced as an independent additive constant.
Experiments \cite{dupaix05} and our recent simulations \cite{hoy06} suggest that $\tau_{flow}$ and $G_R$ scale in the same way and are controlled by the same physical processes.  
For example, both decrease nearly linearly as $T$ increases \cite{vanMelick03,dupaix05}, vanish at a strain-rate dependent $T_g$,
and increase logarithmically with strain rate \cite{wendlandt05}.
Indeed complete strain hardening curves for different rates and cohesion strengths collapsed when scaled by $\tau_{flow}$ \cite{hoy06}.
This suggests that $\tau_{flow}$ is most naturally included as a
multiplicative rather than additive factor. 
To further test this idea we have examined fits to the eight-chain
model for a range of $k_{bend}$, $T$ and strain states.

\begin{figure}[htbp]
\includegraphics[width=3.25in]{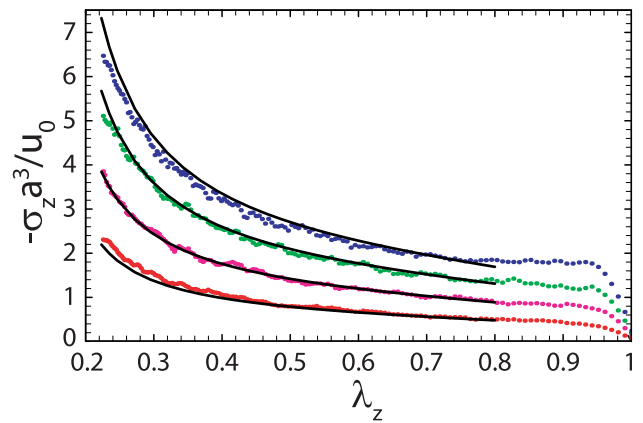}
\caption{ (Color online)
Compressive stress $-\sigma_z$ as a function of $\lambda_z$
for uniaxial compression at $k_B T/u_0 =$ 0, 0.1, 0.2, and 0.3 from top to
bottom.
The chains had $k_{bend}=1.5 u_0$ and $\dot\epsilon=-10^{-4}/\tau_{LJ}$.
Solid lines show a fit to the eight-chain model (Eq.\ \ref{eq:origeightchain})
at $k_B T/u_0 =0.2$ and the extrapolation of this fit to other
temperatures using the modified model (Eq.\ \ref{eq:modifiedeightchain}).
The initial elastic rise and yield for $1 > \lambda_z > 0.8$ are
sensitive to aging and are not fit by the eight-chain model.
}
\label{fig:modifiedeightnofreeuniax}
\end{figure}

Figure \ref{fig:modifiedeightnofreeuniax} shows the compressive stress $-\sigma_{z}$ as a function of $\lambda_z$ for uniaxial compression of systems with
$k_{bend} = 1.5u_0$.
Near $\lambda_z=1$ there is a sharp elastic increase, followed by yield.
Both simulations and
experiments \cite{hasan93,haward97,vanMelick03b,varnik04,rottler05}
find that this initial region ($0.8 \lesssim \lambda_z < 1$) is
sensitive to the past history
of the sample, including the quench rate and aging.
At greater compressions the system is ``rejuvenated'' and the stress
becomes independent of history \cite{hasan93}.
Our discussion will focus on this strain hardening regime.

As with experimental data,
the strain hardening region ($\lambda < 0.8$) of all curves can be
fit (within random stress fluctuations) by adjusting the
parameters ($\tau_{flow}^{zx}$, $G_{R}$, $N_e$) in Equation \ref{eq:origeightchain}.
The quality of such fits is illustrated for $T=0.2 u_0/k_B$.
Typical uncertainties in fit parameters are about 10\%.
For example, data at all temperatures can be fit with $N_e =15 \pm 1$
and best fit values lie within this range.
Note that the value of $N_e=26$ obtained from the plateau modulus
and PPA is substantially larger \cite{everaers04}.
Fits of Eq. \ref{eq:origeightchain} to experiments \cite{arruda93b} also tend to yield smaller values of $N_e$ than the plateau modulus.

As in previous work \cite{hoy06, vanMelick03, dupaix05},
fit values of both the flow stress and hardening modulus decrease
linearly with temperature.
The temperature where they extrapolate to zero,
$T_g \simeq 0.41 u_0/k_B$, is consistent with previous results for the
glass transition temperature for this strain rate \cite{rottler03c}. 
Data for all temperatures can be fit with a fixed ratio
$\beta \equiv G_R/\tau_{flow}$ anywhere in the range from 0.5 to 0.7. 
This nearly constant value of $\beta$ provides further support for the
idea that strain hardening scales with flow stress.

The above observations can be incorporated into a modified eight chain model
that describes the temperature dependent strain hardening
in terms of only four parameters $\tau_{flow}^{0}$, $T_g$, $N_e$
and a temperature independent ratio $\beta$.
Here $\tau_{flow}^0$ is the flow stress at $T=0$ and at other temperatures
$\tau_{flow} = \tau_{flow}^0(1 - T/T_{g})$.
The shear stress in the strain hardening regime is written as
\begin{equation}
\sigma_{i} - \sigma_{j} = \tau_{flow}^{0,ij}\left[1 -  \frac{T}{T_{g}}\right]\left[1 + \beta \displaystyle\frac{L^{-1}(h)}{3h}(\lambda_{i}^{2} -\lambda_j^2)
\right] \ \ .
\label{eq:modifiedeightchain}
\end{equation}
Note that Eq. \ref{eq:modifiedeightchain} is equivalent to the usual
eight-chain model except that it imposes proportionality between
the flow stress and hardening modulus and assumes a linear temperature
drop in both.
This reduction in parameters may be useful in extrapolating experimental
data, since values for one temperature determine those at any other
if $T_g$ is known.

\begin{figure}[htbp]
\includegraphics[width=3.25in]{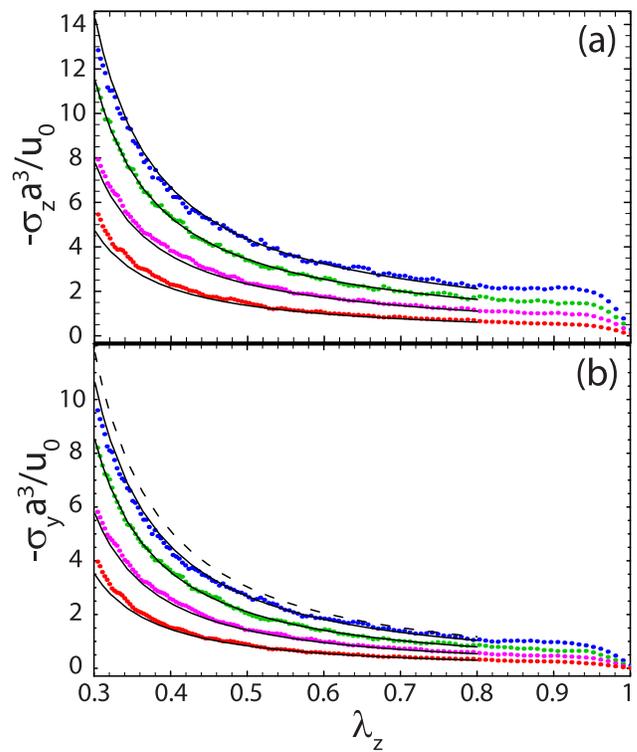}
\caption{
(a) Compressive stress $-\sigma_z$ and (b) perpendicular stress $-\sigma_y$
as a function of $\lambda_z$
for plane strain compression at $k_B T/u_0 =$ 0, 0.1, 0.2, and 0.3 from top to
bottom in each panel.
The chains had $k_{bend}=1.5 u_0$ and $\dot\epsilon=-10^{-4}/\tau_{LJ}$.
Solid lines show a fit to the eight-chain model (Eq.\ \ref{eq:origeightchain})
at $k_B T/u_0 =0.1$ and the extrapolation of this fit to other
temperatures using the modified model (Eq.\ \ref{eq:modifiedeightchain}).
The dashed line in (b) shows the variation in $\sigma_y$ predicted by
Eq.\ \ref{eq:origeightchain} and measured values of $\sigma_z$.
}
\label{fig:modifiedeightnofreeplane}
\end{figure}

The solid lines in Figure \ref{fig:modifiedeightnofreeuniax} show
predictions of Eq.\ \ref{eq:modifiedeightchain}
based on the fit at $T=0.2 u_0/k_B$:
$\tau_{flow}^{0} = 0.634u_0 a^{-3}$, $\beta = 0.56$, and $N_e = 14.1$.
The predictions agree quite well with simulation results over an extremely
wide range of $T/T_g$.
The largest deviations are of order 10\% at $T=0$ and the smallest
$\lambda_z$.
There is a slight over-prediction of the change in curvature with
increasing $T$ that manifests as slight ($\sim 10$\%) increases
in $\beta$ or decreases in $N_e$ in best fits to the data, particularly
at $T=0.3$.
Fits of the same quality are obtained for all $k_{bend}$ and strain
states (see Fig. \ref{fig:modifiedeightnofreeplane}), suggesting
that this simple extrapolation may be widely applicable to
data from simulations or experiment. 

A more stringent test of Eq.\ \ref{eq:modifiedeightchain} is whether it is
able to predict stresses for multiple strain states with the same
parameters.
As pointed out by Arruda and Boyce \cite{arruda93b}, uniaxial and plane
strain compression produce extremely different changes in chain
configuration.
Under plane strain compression the chains all stretch in
one direction, while in uniaxial compression the chains stretch along all
directions in the plane perpendicular to the compression axis.

Figure \ref{fig:modifiedeightnofreeplane} shows results for plane strain
compression of the same systems as Fig.\ \ref{fig:modifiedeightnofreeuniax}.
As before, excellent fits can be obtained to Eq. \ref{eq:origeightchain}
and extrapolations from fits at one temperature using Eq. \ref{eq:modifiedeightchain} capture the variation
in stress over the full temperature range.
Despite these successes,
there are troubling inconsistencies in the parameters of these
fits.
While the values of $G_R$ for $\sigma_{z}$ in plane strain and uniaxial
compression are consistent (within our 10\% uncertainty),
the value of $N_e$ is significantly higher for plane strain; $N_e=20\pm 2$.
In addition, the strain hardening of $\sigma_{y}$ and $\sigma_{z}$
in plane strain are inconsistent.  From Eq. \ref{eq:origeightchain},
the value of $\sigma_{y}$ is
determined up to an additive constant from measurements of $\sigma_{z}$.
The dashed line in Fig. \ref{fig:modifiedeightnofreeplane}(b) shows
this prediction for $T=0$ with the additive constant adjusted to fit
data at large $\lambda_z$.
At all temperatures the variation in $\sigma_{y}$ is systematically
smaller than predicted from $\sigma_{z}$.
The decrease corresponds to a reduction in $G_R$ for $\sigma_y$ by
about 20\%.
In contrast, best fit values of $N_e$ for $\sigma_y$ and $\sigma_z$
match to within 1\%. 

A strong dependence of model parameters such as $G_R$ and $N_e$ on strain state has also been noted experimentally \cite{gsell94}. 
One possibility is that the strain state changes the role of entanglements
in ways that are not captured completely by the eight-chain model.
Another is that while the eight-chain model provides a useful fitting
function, it does not capture the correct strain hardening mechanisms.
In this case, fit parameters may not have direct physical significance.
The scaling of $G_R$ with $\sigma_{flow}$ rather than $T$ supports
the view that entropy is not the dominant source of stress and this
is examined further below.
It is also important
to note that the fits above assumed
constant volume, but the simulation volume decreased with strain by
up to 8\% for plane strain and large $k_{bend}$.
Including the correct $\lambda_i$ in Eq. \ref{eq:origeightchain} would
change the predicted stress by up to 15\%.
In the following section we show that violations of the assumption of
affine displacement of entanglements can produce similar changes.
Thus the fit parameters compensate for changes that are not included
in the model, further reducing their direct relevance.

\subsection{Dissipative and Energetic Stresses}
\label{subsec:disseng}

Entropic network models assume that strain hardening arises entirely from a reversible increase in the  entropy of the entanglement network.
Experiments showed many years ago that strain hardening is also associated with large increases in internal energy \cite{hasan93}, but this observation has not been incorporated into published theoretical models.
Simulations allow us to separate the role of entropy and energy in strain
hardening.
 
For uniaxial and plane strain compression,
the stress along the compressive axis is directly related to the work $W$
done on the system per unit strain:
$\sigma_z = \partial W /\partial \epsilon_z$.
$\sigma_z$ can be separated into an energetic component $\sigma_z^U$
and a thermal component $\sigma_z^Q$
using the first law of thermodynamics:
$dW = dQ + dU$, where $U$ is the internal energy of the system 
and $dQ$ is the heat transfer \textit{away from} the system.
This implies
\begin{equation}
\begin{array}{ccccc}
\sigma_z = \displaystyle\frac{\partial W}{\partial \epsilon_z}
& : & \sigma_z^{U} = \displaystyle\frac{\partial U}{\partial \epsilon_z}
& : & \sigma_z^{Q} = \displaystyle\frac{\partial Q}{\partial \epsilon_z}
= \sigma_z - \sigma_z^{U}.
\end{array}
\end{equation}
These quantities are readily obtained from our simulation data
and could in principle be obtained by differentiating results for $W$ and $Q$
from deformation calorimetry experiments.
Unfortunately existing studies \cite{adams88, salamatina94, oleinik06} have not extended into the strain
hardening regime.

Experimental data are frequently plotted in a manner designed
to isolate the Gaussian and Langevin
contributions to the strain (Eq. \ref{eq:langevinmodel}).
If $\sigma_z$ is plotted against $g(\bar\lambda) = \lambda_z^2-\lambda_x^2$,
then Eq. \ref{eq:origeightchain} predicts a straight line
in the Gaussian limit ($h << 1$).
The Langevin correction ($(3h)^{-1}L^{-1}(h)$)
adds an upwards curvature.
Since $\lambda_x$ is not generally measured, experimental stresses are
plotted as a function $g(\lambda_z)$ that is determined
by assuming constant volume.
For uniaxial strain $g(\lambda_z) = \lambda_z^2-1/\lambda_z$
and for plane strain $g(\lambda_z) = \lambda_z^2-1/\lambda_z^2$.

\begin{figure}[htbp]
\includegraphics[width=3.25in]{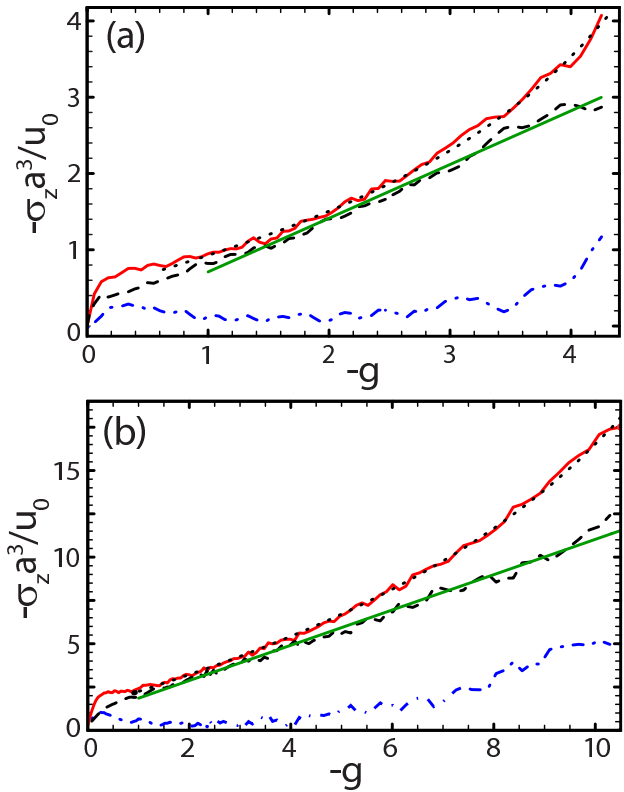}
\caption{(Color online) Total stress (solid lines) and thermal (dashed lines)
and potential energy (dot-dashed lines) contributions
for (a) uniaxial compression at $k_B T/u_0 = 0.2$ and
(b) plane strain compression at $k_B T/u_0 = 0$.
The systems had $k_{bend}=2.0$ ($N_e=22$), $N=350$ and
(a) $\dot\epsilon=-10^{-5}/\tau_{LJ}$ or (b) $\dot\epsilon=-10^{-4}/\tau_{LJ}$.
Dotted lines show best fits of $\sigma_z$ to Eq.\ \ref{eq:modifiedeightchain}
with (a) $N_e=15.5$ and (b) $N_e = 22$.
Straight lines are fits to $\sigma_z^Q$.
Both $\sigma_z$ and $g$ are negative under compression.}
\label{fig:threestresses}
\end{figure}

Figure \ref{fig:threestresses} illustrates the variation of total,
thermal and energetic stresses with $g$ under uniaxial and plane strain
compression for the most highly entangled system ($N_e=22$).
The total stress for both strain states
shows strong upward curvature that is normally
attributed to the Langevin correction.
As expected from entropic network models,
the amount of curvature decreases with increasing $N_e$ \cite{hoy06}. 
However, Fig. \ref{fig:threestresses} shows that this curvature
is not related to entropy.
Almost all of the upward curvature is associated with the energetic
contribution to the stress, while $\sigma_z^Q$ shows the
linear behavior expected for Gaussian chains.
Similar behavior is observed for all $N_e$ and $T$,
and we now discuss the trends in $\sigma^Q$ and $\sigma^U$ separately.

For all systems, temperatures, and strain states the thermal stress
is well fit by the linear behavior expected for Gaussian chains.
There may be a small upwards curvature, particularly for
uniaxial compression, but it is comparable to statistical fluctuations.
Attempts to fit $\sigma_z^Q$ to the eight chain model always require
increasing $N_e$ significantly above values obtained from the
melt plateau modulus.

We define a thermal hardening modulus $G_{therm}$ from the slope
of linear fits to
$\sigma_z^{Q} = \sigma_{0} + G_{therm}g(\lambda_z)$.
Table \ref{tab:Gplast} shows values for $G_{therm}$ for uniaxial and
plane strain compression for various entanglement lengths at $T=0.2u_0/k_B$.  
While $G_{therm}$ is systematically higher for uniaxial compression,
the differences (10-30\%) are not large.
In contrast, $G_{therm}$ decreases rapidly with increasing $N_e$.
Rubber elasticity theories for Gaussian chains would predict
$G_R \propto N_e^{-1}$ and it is interesting that $G_{therm}$
appears to scale in this way.
As shown in Table \ref{tab:Gplast}, changes
in $N_{e}G_{therm}$ are within our statistical
error bars ($\sim 10$\%) and show no systematic trend with $N_e$.

\begin{table}[h]
\caption{Thermal Moduli $G_{therm}$ in units of $u_0/a^3$ for
$T=0.2u_0/k_B$ and $\dot\epsilon=-10^{-4}/\tau_{LJ}$.
Errorbars are about 10\%.}
\begin{ruledtabular}
\begin{tabular}{lcccc}
$N_{e}$ & $G_{therm}^{uniax}$ & $G_{therm}^{plane}$ & $N_{e}G_{therm}^{uniax}$ & $N_{e}G_{therm}^{plane}$\\
22 & 1.3 & 1.0 & 28 & 23\\
26 & 1.0 & 0.75 & 27 & 20\\
39 & 0.57 & 0.50 & 22 & 20\\
71 & 0.37 & 0.34 & 26 & 24
\end{tabular}
\end{ruledtabular}
\label{tab:Gplast}
\end{table}

\begin{figure}[h!]
\includegraphics[width=3.25in]{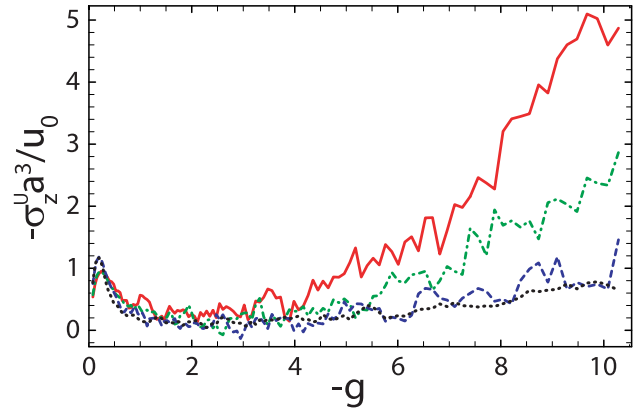}
\caption{(Color online) Energetic components of stress $\sigma_{z}^{U}$ for plane strain compression at $T=0$ with strain rate $\dot{\epsilon} = -10^{-4}/\tau_{LJ}$ for $N_{e}=22$ (solid line), $N_{e}=26$ (dot-dashed line), $N_{e}=39$ (dashed line), and $N_e=71$ (dotted line).}
\label{fig:sigmaengdiffNe}
\end{figure}

Figure \ref{fig:sigmaengdiffNe} shows $\sigma_z^{U}$ during plane strain
compression for different $N_e$.
Results are shown for $T = 0$ because the energetic stresses are the largest,
but results at higher temperatures show similar trends.
The value of $\sigma^{U}$ rises to a peak near the yield point,
and then drops to a nearly constant value for $|g|> 1 $.
The initial behavior for $|g| <1$ is nearly independent of the
entanglement length 
but does depend weakly on age and strain rate.
The behavior at slightly larger $|g|$ depends strongly
on entanglement length (Fig. \ref{fig:sigmaengdiffNe}).
For example, at $T=0.2u_0/k_B$ the ratio of the constant energetic
stress to the flow stress rises from about 4\% for flexible chains
to 16\% for the most entangled system.
Similar ratios are obtained for the $T=0$ data in
Fig. \ref{fig:sigmaengdiffNe}.

Hasan and Boyce examined the enthalpy stored in samples of
polystyrene (PS), polymethylmethacrylate (PMMA) and polycarbonate (PC)
as a function of residual strain \cite{hasan93}.
They found a sharp increase in enthalpy up to a strain of about
20\% ($|g| = 0.55$) that was larger in annealed samples than
in rapidly quenched samples like those used here.
For quenched PS, the magnitude of the rise in energy density
is about 4MPa.
Values for the work performed are difficult to extract from the
paper, but as a rough estimate we take the flow stress (55MPa)
times the strain (0.2) and find 11MPa.
Thus in the initial stages of deformation,
of order a third of the work is stored in energy.
Calculating the same ratio for our simulations gives
values between 30 and 45\%.

For strains from -0.2 to -0.8 ($|g|=2.0$) or larger,
Hasan and Boyce found a weak, nearly linear rise in enthalpy.
This corresponds to a constant $\sigma^U$ like that observed
in Fig. \ref{fig:sigmaengdiffNe} for intermediate $|g|$.
Analysis of their figures \cite{hasan93} shows that the
ratio of $\sigma^U$ to the flow stress increases from
about 4\% for PS to 15\% for PC.
Since PC is more entangled than PS, this trend is the same
as observed in Fig. \ref{fig:sigmaengdiffNe}.
Note that in both simulations and experiments the fraction of
work stored as energy depends strongly on the strain amplitude.
The fraction stored during the initial rise to the flow stress
is dependent on sample age \cite{hasan93} and may be of order
50\% or more \cite{oleynik89,shenogin02}. 
As $|g|$ increases and $\sigma^U$ remains constant,
the fraction stored as energy
drops towards the much smaller value
given by $\sigma^U/\sigma_{flow}$.

Fig. \ref{fig:sigmaengdiffNe} shows a sharp rise
in $\sigma^U$ at the largest values of $|g|$.
The onset moves to smaller $|g|$ and the magnitude of the rise
increases as $N_e$ decreases.
This rise is the source of almost all the curvature in the total stress.
The experiments of Hasan and Boyce \cite{hasan93} did not show this rise.
One reason may be that their experiments only extended to uniaxial strains
of -0.8 ($|g|=2.0$) for the most entangled systems.
However, their measurement is also limited to the residual enthalpy
after the sample is unloaded.
Simulations were performed to determine the amount of energy
recovered during unloading from different strains.
The recovered energy is relatively small in the region where
$\sigma^U$ has a small constant value, but rises rapidly
at larger strains.
Indeed, almost all of the energy that contributes to the sharp rise
in $\sigma^U$ at large $|g|$ is recovered when samples
are unloaded.
Thus experiments could only observe this rise by using
deformation calorimetry.

The rise in energetic stress at large $|g|$ seems to occur
when segments of length $N_e$ are pulled taut between entanglements.
To demonstrate this we examined changes in chain statistics
with stretch.
Entropic network models assume that the deformation of the
entanglement network is affine to the macroscopic stretch.
An affine displacement would, on average, increase the
length of any chain segment by a factor of $\lambda_{chain}$.
This stretch can not apply at the smallest scales since
the length of chemical bonds $l_0$ can not increase significantly.
As a result chains are pulled taut and deform subaffinely on small scales.
The larger the strain, the larger the length of taut segments.

To illustrate this we compare the rms Euclidean distance
between monomers
separated by n bonds $R(n)$ to the affine prediction
$R_{aff}(n)$.
If $R_0(n)$ is the distance before stretching,
then $R_{aff}(n) = \lambda_{chain} R_0(n)$ where
$\lambda_{chain}^2 = (\lambda_x^2+\lambda_y^2+\lambda_z^2)/3$.
For the case of plane strain,
$\lambda_{chain}^2=(\lambda_z^2 + 1 + \lambda_z^{-2}(V/V_{0})^{2})/3$,
where $\lambda_x$ has been eliminated by using
the ratio $V/V_0$ of the final and initial volumes.
Volume changes are normally ignored, but are large enough to affect
the plots shown below.

Figure \ref{fig:affinityenergy}(a) shows the ratio of the observed $R(n)$
to the affine prediction as a function of $n/N_e$ for different
$g(\lambda_z)$ and $N_e$ under plane strain. 
There is a clear crossover from subaffine behavior ($R/R_{aff} <1$)
to affine behavior ($R/R_{aff} \simeq 1$) with increasing $n$.
In the subaffine regime at small $n$, chains are pulled nearly
taut.
The crossover to affine behavior moves to larger $n$ as $|g|$
increases, implying that chains are pulled straight over longer
segments.
For chains with $N_e=39$ the crossover remains slightly below $N_e$ at
the largest strains considered here.
However for $N_e=22$ the crossover appears to reach $N_e$
by $|g|=5$.
At larger $|g|$ the magnitude of $R/R_{aff}$ decreases, but
the region of rapid crossover appears to remain near $N_e$.
This suggests that the entanglements prevent chains from stretching
taut on longer scales.

Figure \ref{fig:affinityenergy}(b) shows there is a direct correlation
between subaffine deformation at $N_e$ and the increase in the energetic
contribution to the stress.
The values of $\sigma_z^U$ from Fig. \ref{fig:sigmaengdiffNe}
are replotted against $R(N_e)/R_{aff}(N_e)$ instead of $|g|$ \cite{foot7}.
There is a sharp rise in $\sigma_z^U$ as $R(N_e)/R_{aff}(N_e)$
decreases below about 0.925.
As seen in panel (a), this corresponds roughly to the point
where the length of taut segments reaches $N_e$.
This suggests that the energetic stress arises when the entanglement
network begins to resist further deformation.
As expected from this picture, we find a growing tension in
covalent bonds as $\sigma_z^U$ rises.
However, the maximum tensions in the ``worst case'' scenario of plane strain compression for $N_e = 22$ are only about $100u_0/a$, which is well below the breaking strength $240u_0/a$ used in breakable-bond simulations \cite{rottler03}. 

The above findings help to explain some of the discrepancies
in the fit parameters for the eight-chain model.
The upwards curvature in plots of $\sigma_z$ vs. $|g|$
comes from energetic terms rather than entropy.
Fits to uniaxial and plane strain give different $N_e$
because the energetic contributions are different.
However the fit values are never far from $N_e$ because
the sharp increase in $\sigma_z^U$ occurs when segments
of length $N_e$ are pulled taut.
Fig. \ref{fig:affinityenergy}(a) also indicates that the
entanglement network does not deform completely affinely
as assumed in the eight-chain model.
Even at small strains $R(N_e)/R_{aff}(N_e)$ is slightly
less than one and this would produce significant
($\sim$10\%) changes in the stress from Eqs. \ref{eq:origeightchain}
or \ref{eq:modifiedeightchain} \cite{foot3}.

Note that the deviations from affinity observed in Fig. \ref{fig:affinityenergy}
are significantly smaller than those predicted from rubber-elasticity
based models for the non-affine deformations in entangled polymers
above $T_g$ \cite{warner78,read07}.
These models assume that fluctuations about affine deformation are
confined to the ``tube'' formed by surrounding entanglements.
They predict nonaffine reductions in $R(N_e)/R_{aff}(N_e)$ that are
about 50\% greater than our results at small strains.
At the larger strains where we find entanglements produce a significant
energetic stress, the disparity decreases.
The discrepancy appears to reflect the fact that the
nonaffine displacements in our simulations \cite{hoy06}
are much smaller than the tube radius until the energetic stress begins
to dominate.
Interestingly, the magnitude of the
nonaffine displacements decreases slightly with increasing $r_{c}$,
suggesting they are limited by cohesive interchain interactions rather
than entanglements.
These observations provide further evidence that polymers in a glass are
not free to explore their tube as assumed in entropic models.
It would be interesting to extend these comparisons to melt models to
see what additional information can be obtained.

\begin{figure}[htbp]
\includegraphics[width=3.25in]{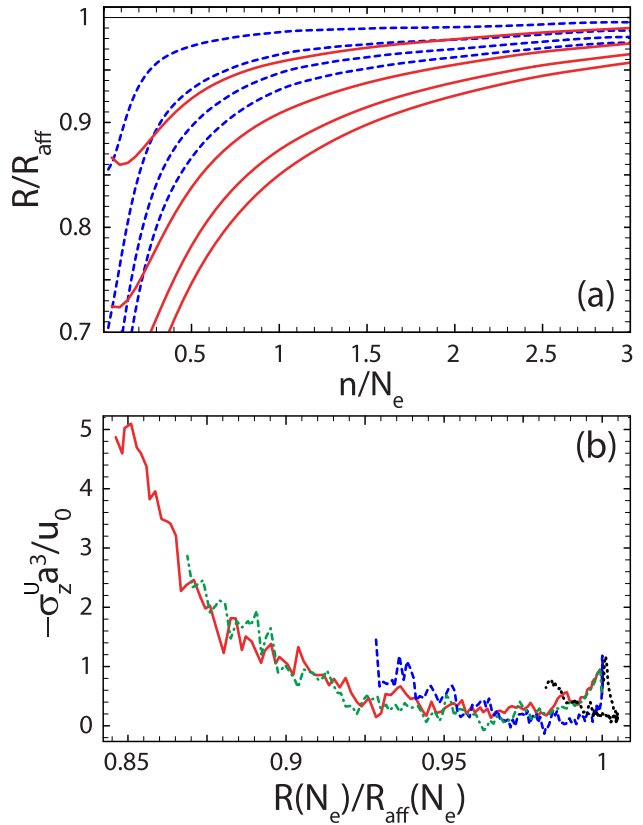}
\caption{(Color online)
(a) $R(n)/R_{aff}(n)$ for the same systems and conditions as
Fig. \ref{fig:sigmaengdiffNe} with $n$ scaled by $N_e$.
Solid lines indicate $N_e=22$ and dashed lines indicate $N_e=39$.
Curves are for $|g|=$2.5, 5, 7.5, and 10 from top to bottom.
(b) $\sigma_{z}^{U}$  plotted against $R/R_{aff}$ evaluated at $n=N_e$.
This corresponds to evaluating $\sigma_{z}^{U}$ along a vertical slice of (a).
Solid, dot-dashed, dashed, and dotted lines indicate data for $N_e=$22, 26, 39,
and 71 respectively \cite{foot7}.}
\label{fig:affinityenergy}
\end{figure}

\subsection{Reversibility and entropic back stresses}
\label{subsec:reverse}

If the work performed in deforming a glass is entropic, it should be reversible.
However, large strain experiments performed well below $T_g$
show only $\sim$10\% strain recovery upon unloading \cite{hasan93}.
Entropic network models postulate that there is an entropic ``back stress''
\cite{arruda93b} that favors further strain reduction but that
relaxation is too slow to observe because of
the high viscosity of the glassy state \cite{haward68}.
As expected from this picture, lowering the viscosity by heating 
even slightly above $T_{g}$ leads to nearly complete shape recovery
of well-entangled glasses \cite{haward97}.

To see if similar behavior occurs in simulations, we loaded 
samples to $\epsilon = -1.5$ at $\dot\epsilon = -10^{-5}/\tau_{LJ}$
and $T=0.2u_0/k_B$.
The samples were then unloaded at the 
same $|\dot\epsilon|$ and $T$ until all $\sigma_i$ were zero.
Fully entangled samples ($N=350$, $N_e=39$) recovered only
6\% of the peak strain and unentangled chains ($N=16$)
recovered slightly less, $\sim 4$\%.
The result for entangled chains is comparable to experiments \cite{hasan93}.

\begin{figure}[htbp]
\includegraphics[width=3.25in]{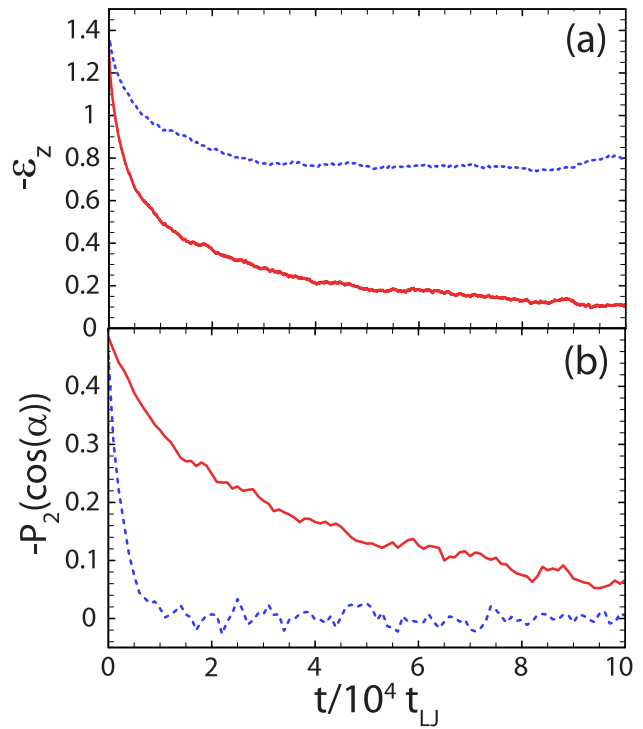}
\caption{(Color online) Time dependent relaxation at $T=0.4 u_0/k_B$
of (a) true strain and (b) chain
orientation parameter $P_2$ for entangled $N=350$ (solid lines)
and unentangled $N=16$ (dashed lines) systems.
Systems were prepared by loading to $\epsilon_z=1.5$ at $T=0.2u_0/k_B$,
unloading to zero stress and then heating to 0.4$u_0/k_B$ over 100$\tau_{LJ}$.
}
\label{fig:shaperecovery}
\end{figure}

The samples were then heated to $T=0.4u_0/k_B$ over $100 \tau_{LJ}$
and allowed to relax
with a Nose-Hoover barostat imposing zero stress in all three directions.
Figure \ref{fig:shaperecovery}(a) shows the resulting strain recovery.
For the entangled system,
an additional 87\% of the strain was recovered after $10^{5}\tau_{LJ}$
and the rate of recovery remained significant at the end of this period.
In the unentangled system, 46\% of the strain is recovered,
mainly in the first $2 \times 10^4 \tau_{LJ}$.
While this recovery is substantially smaller than for entangled systems,
network models would predict no strain recovery for unentangled chains.
Examination of pair and bond energies shows that they are nearly
constant during the relaxation at $T=0.4u_0/k_B$.
These results imply that entropic stresses drive the relaxation and
that entanglements play an important role.

To monitor the entropy in chain confirmations we evaluated 
the orientational order parameter,
$P_{2}(cos(\alpha)) = (3 cos^{2}(\alpha) - 1)/2$,
where $cos^{2}(\alpha) = <R_{z}^{2}>/<R_{ee}^{2}>$.
This quantity measures the deviation from isotropy at the end-end scale.
There is significant orientation of both short and long chains during
the initial strain, which is discussed further below.
During the relaxation to zero stress and heating to $T=0.4u_0/k_B$
the orientation relaxes only 3\% for entangled chains and 5\%
for unentangled chains.
As shown in Fig. \ref{fig:shaperecovery}(b), 
rapid and substantial deorientation occurs during the strain relaxation
above $T_g$.
Short chains become nearly isotropic after only $\sim 2 \cdot 10^{4}\tau_{LJ}$, 
and there is little strain recovery after this point.
Entangled chains deorient more slowly, and both $P_2$ and $\epsilon_z$ are
continuing to evolve slowly at the end of the simulations.
These results clearly show that the entropy of chain orientation drives the  
strain relaxation.
They also show that the network of entangled chains prevents chains from
deorienting without recovery of the macroscopic strain.

While Fig. \ref{fig:shaperecovery} provides strong support
for an entropic back stress, the magnitude of this stress
can only be of order $\rho_e k_B T_g$ and thus much
smaller than the stresses associated with strain hardening.
To confirm this we took the $N=350$ sample studied in
Fig. \ref{fig:shaperecovery}
and heated to $T=0.4u_0/k_B$ with different stress control.
Instead of fixing all $\sigma_i$ to zero, only the total pressure
$p = -(\sigma_x+\sigma_y+\sigma_z)/3$ was kept at zero while
the ratios of the $L_i$ were fixed at the values after deformation.
After heating, there was a shear stress
$\sigma_z-\sigma_x \approx \sigma_z-\sigma_y$ whose
direction favored relaxation back to $\epsilon=0$.
The magnitude of this stress relaxed rapidly ($\sim 10^4\tau_{LJ}$) to about
twice the entropic estimate of $\rho_e k_B T_g \approx 0.02u_0/a^3$,
while the stress during strain hardening below $T_g$ is more than two
orders of magnitude larger.
Similar results were obtained at higher temperatures.
We next applied a shear force of the same magnitude ($0.04u_0/a^3$) to
an unstrained system at $T=0.4u_0/k_B$.
The magnitude of the strain produced by this stress over
$10^5\tau_{LJ}$ was $1.2$, which is
comparable to that during stress relaxation (Fig.  \ref{fig:shaperecovery}).
Both this driven response and the stress relaxation varied
approximately as the logarithm of time, indicating that
the sample displays creep rather than viscous flow.
Recent studies of a similar glassy system also show
creep behavior at this temperature and time scale \cite{warren07}.

\subsection{Chain Length Effects}
\label{subsec:chainlengthdep}

The orientation of unentangled chains shown in Fig. \ref{fig:shaperecovery}(b)
is not expected from entropic network models.
For $N < N_e$ there is no entanglement network spanning the system.
Network models assume that this network is essential in forcing
the deformation of individual chains to follow the
macroscopic strain.
However because chains are not free to relax in the glassy state,
chain orientation can occur even without entanglements.
In recent work on uniaxial compression, we found significant strain
hardening of unentangled
chains \cite{hoy06} and discovered a direct connection to chain
orientation \cite{hoy07} as suggested by recent analytic studies
\cite{ken}.
In this section we extend the study of chain length dependence to other
strain states and systems.

\begin{figure}[htbp]
\includegraphics[width=3.25in]{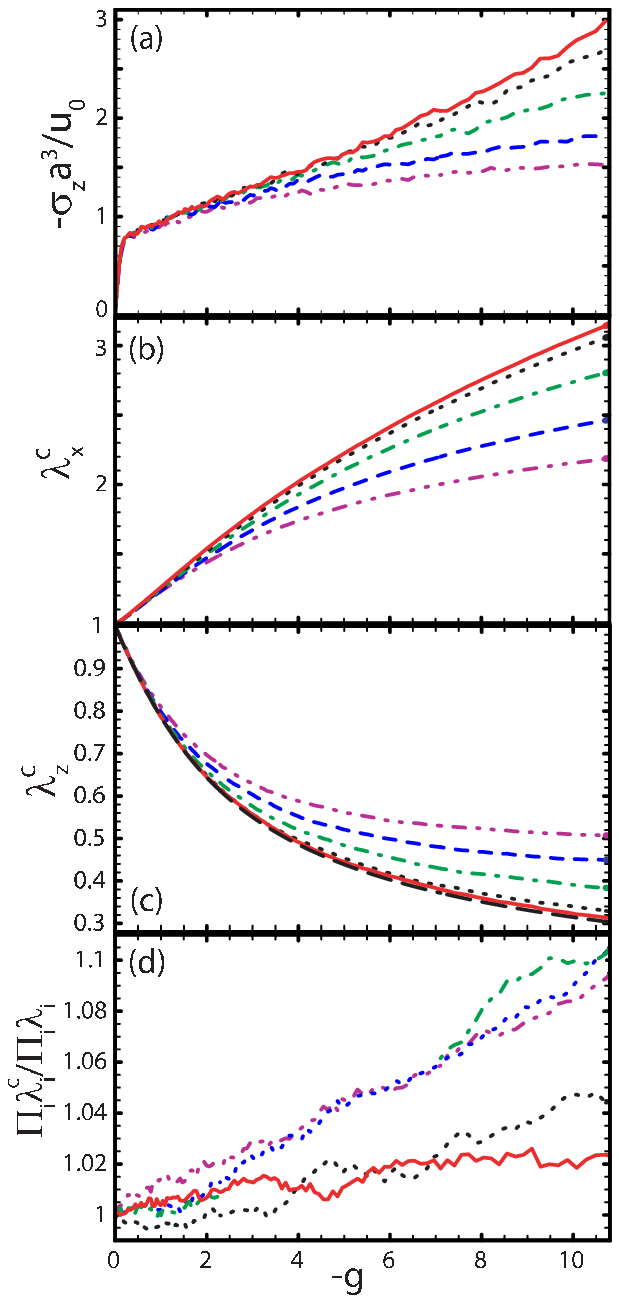}
\caption{(Color online)
Dependence of (a) stress and (b,c) microscopic chain orientation
on $|g|$ for flexible chains under plane strain compression
at $T=0.2u_0/k_B$ with $\dot\epsilon=-10^{-5}/\tau_{LJ}$.
Panel (d) shows the ratio of changes in chain volume to
macroscopic volume.
The chains have lengths $N = 500$ (solid lines), $N=107$ (dotted lines),
$N=36$ (dash-dotted lines), $N=18$ (dashed lines), and $N=12$ (dash-dot-dotted lines).
The lower dashed line in (c) shows the macroscopic stretch $\lambda_z$.
}
\label{fig:Ne71}
\end{figure}

Figure \ref{fig:Ne71}(a) shows stress-strain curves for flexible chains
($N_{e} = 71$) in plane strain compression for a range of $N$ between
12 and 500. 
At small $|g|$, the stresses are nearly independent of $N$.
Beyond yield, the stresses increase faster for larger $N$, reaching an
asymptotic limit for $N\gg N_e$ as expected from network models \cite{hoy06}.
However, there is significant strain hardening for chains as short as $N_e/6$.
Similar behavior is observed for all entanglement densities under
both uniaxial and plane strain.
This is illustrated for $k_{bend}=2.0u_0$ ($N_e=22$) under uniaxial strain in
Fig. \ref{fig:Ne22}(a)
and for $k_{bend}=0.75u_0$ in Fig. 2(c) of Ref. \cite{hoy07}.

Examination of individual chain conformations shows that strain
hardening is directly correlated with increasing chain
orientation \cite{argon73}.
To quantify this we define a microscopic stretch of chains as
$\lambda_i^c \equiv  R_i/R_i^0$ where $R_i$ is the rms
projection along $i$ of the end-end distance and $R_i^0$ is
the value before deformation.
Fig. \ref{fig:Ne71} shows $\lambda_i^c$ for flexible chains under
plane strain compression.
For fully entangled chains ($N/N_e \sim 7$) the
microscopic stretch remains close to the macroscopic stretch
as already concluded from  Fig.  \ref{fig:affinityenergy}.
For $\lambda_x$ the deviation is smaller than the line width.
For $\lambda_z$ the maximum deviations of about 2\%
(lowest dashed line in Fig. \ref{fig:Ne71}(c)) 
can be attributed to non-affine
deformation of the unentangled ends of the chains.

As the chain length decreases, the microscopic stretch shows increasing
deviations from the macroscopic stretch.
Chains compress by less than the imposed strain along the $z$ axis,
and stretch by a smaller amount along the $x$ axis.
For each $N$, $\lambda_i^c$ is close to the entangled results at
small $|g|$ and then saturates at large $|g|$.
The onset of saturation in $\lambda_i^c$
correlates with the saturation of the stress,
and moves to larger $|g|$ with increasing $N$.
These results clearly show that entanglements force the chain orientation
to follow the macroscopic stretch but that significant
chain orientation occurs without entanglements.
Strain also orients chains in unentangled melts, but is only appreciable
when the strain rate is faster than chain relaxation times \cite{doi86}.
The extremely slow dynamics in glasses prevents relaxation of
shear-induced orientation.

Fig. \ref{fig:Ne71}(d) shows the ratio of the product of the
chain and macroscopic stretches $\Pi_i \lambda_i^c/\Pi_i \lambda_i$.
This corresponds to the ratio of changes in the volume
subtended by the chains to changes in the macroscopic volume.
For entangled chains the ratio is close to unity,
as expected for a crosslinked network.
The volume subtended by unentangled chains need not follow
the macroscopic volume, but the observed deviations are less
than 11\% in Fig. \ref{fig:Ne71}(d).
Deviations are even smaller for flexible chains under uniaxial strain.

\begin{figure}[htbp]
\includegraphics[width=3.25in]{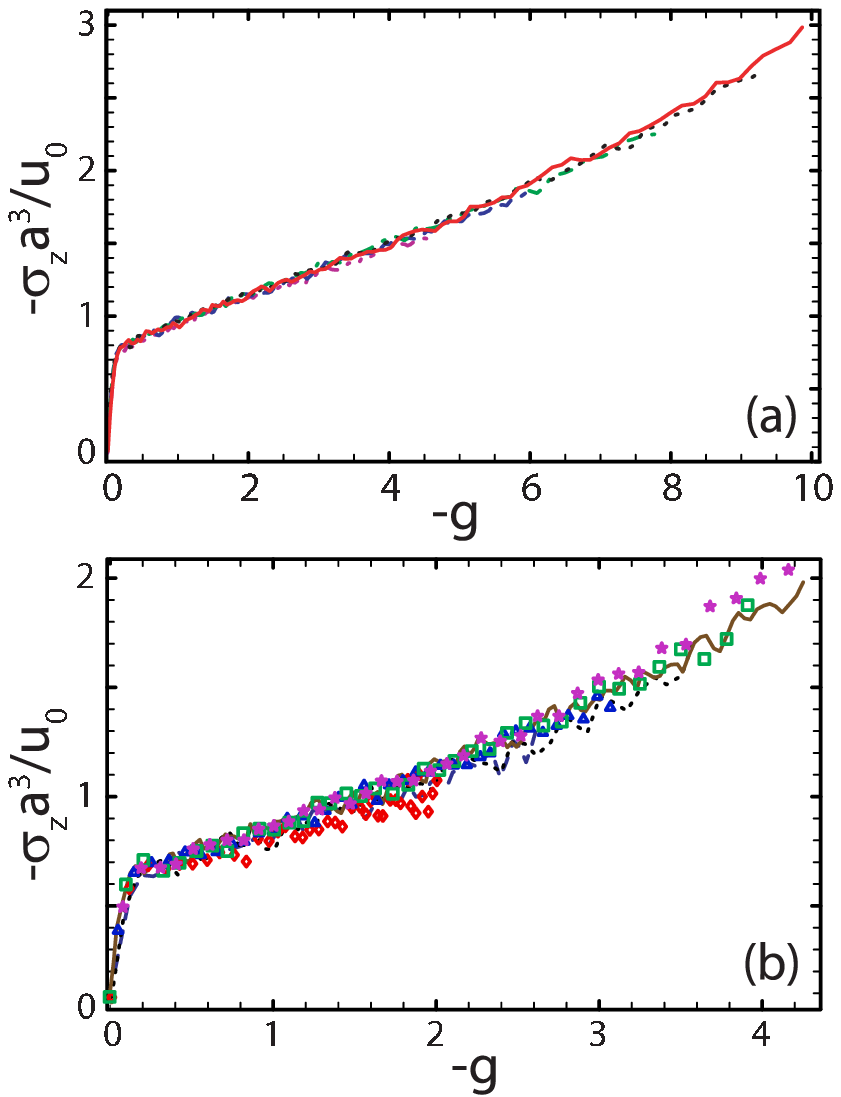}
\caption{
(Color online)
Compressive stress as a function of microscopic orientation function
$g_{eff} \equiv (\lambda_z^c)^2-(\lambda_x^c)^2$ at
$T=0.2u_0k_B$ with $\dot\epsilon=-10^{-5}/\tau_{LJ}$.
(a) Plane strain compression of flexible chains with
lengths $N = 500$ (solid line), 107 (dotted line),
$36$ (dash-dotted line), $18$ (dashed line), or $12$ (dot-dash-dot line).
(b) Uniaxial compression of semiflexible chains ($N_e=39$) with
$N=350$ ($\star$), 175 ({\Huge -}), 70 (squares), 40 ($\cdots$),
25 ($\triangle$), 16 ($- - -$) and 10 ($\diamond$).
}  
\label{fig:collapse}
\end{figure}

Figure \ref{fig:collapse} shows that $\sigma_z$ is determined
directly by the microscopic orientation of chains rather
than the macroscopic deformation.
Results for plane strain compression of flexible chains
(from Fig. \ref{fig:Ne71}(a)) and
uniaxial compression of semiflexible chains ($N_e=39$)
are plotted against an effective $g$ calculated from $\lambda^c_i$:
$g_{eff} \equiv (\lambda_z^c)^2-(\lambda_x^c)^2$.
When plotted against this measure of microscopic chain
orientation, results for all chain lengths collapse onto a universal
curve.
A similar collapse was obtained in Ref. \cite{hoy07} using a
single effective orientation parameter $\lambda_z^{eff}$ along the
compression direction.
This was obtained by measuring $\lambda_x^c$ and using the
assumption of constant chain volume to determine $\lambda_z^{eff}$
(i.e. $\lambda_z^{eff}=1/\lambda_x^c$ for plane strain).
The collapse produced for $g(\lambda_z^{eff})$ is nearly
identical to that in Fig. \ref{fig:collapse}
because chain volume is nearly constant (Fig. \ref{fig:Ne71}(d))
and $g_{eff}$ is mainly determined by $\lambda_x$.

\begin{figure}[htbp]
\includegraphics[width=3.25in]{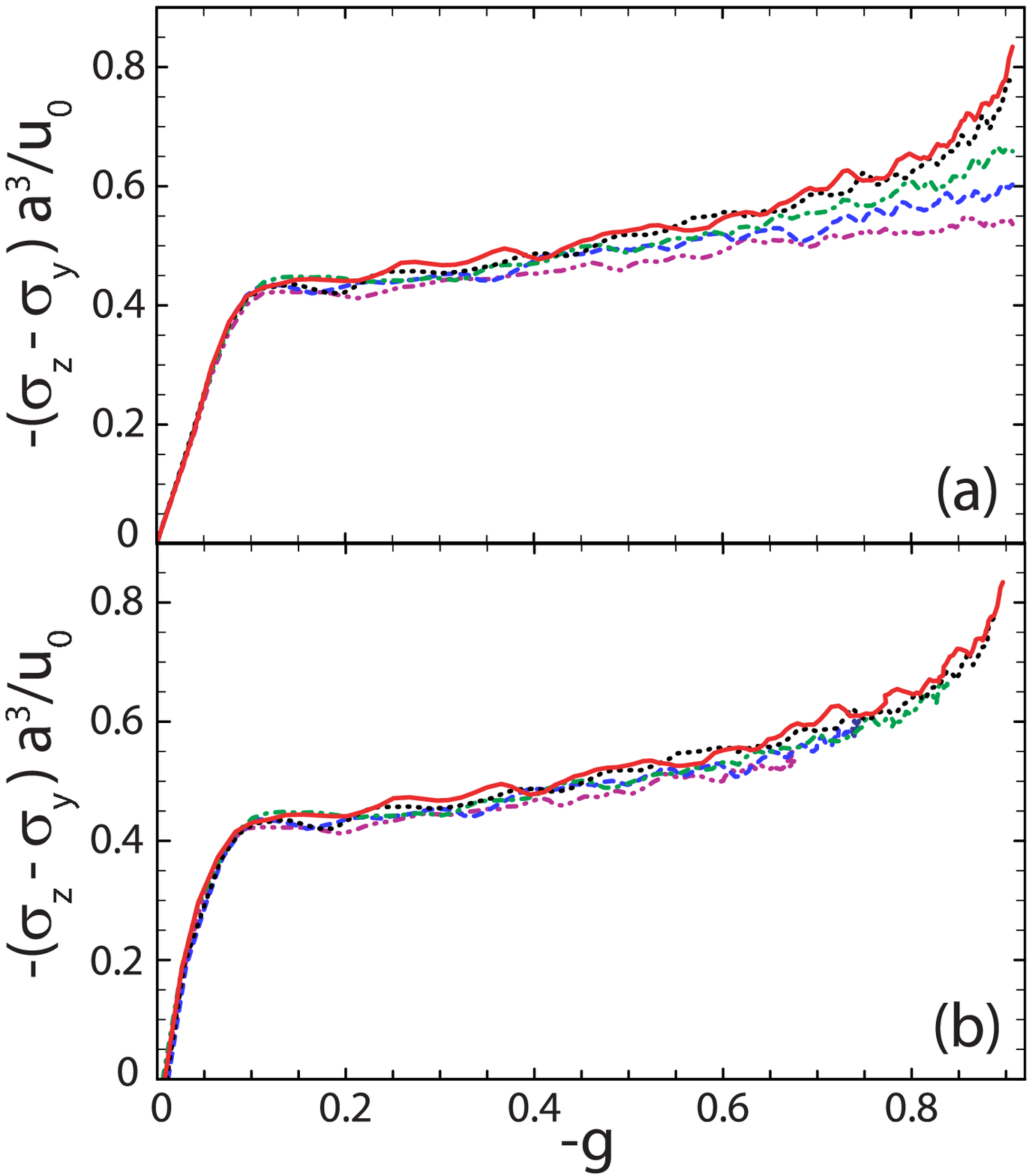}
\caption{(Color online)
Stress difference $\sigma_z-\sigma_y$ as a function of
(a) the macroscopic $g$ and (b) the microscopic orientation function
$g_{eff} \equiv (\lambda_z^c)^2-(\lambda_y^c)^2$ at
$T=0.2u_0k_B$ with $\dot\epsilon=-10^{-5}/\tau_{LJ}$.
Chains have length $N = 500$ (solid line), 107 (dotted line),
$36$ (dash-dotted line), $18$ (dashed line), or $12$ (dot-dash-dot line).
}  
\label{fig:collapse2}
\end{figure}

\begin{figure}[htbp]
\includegraphics[width=3.25in]{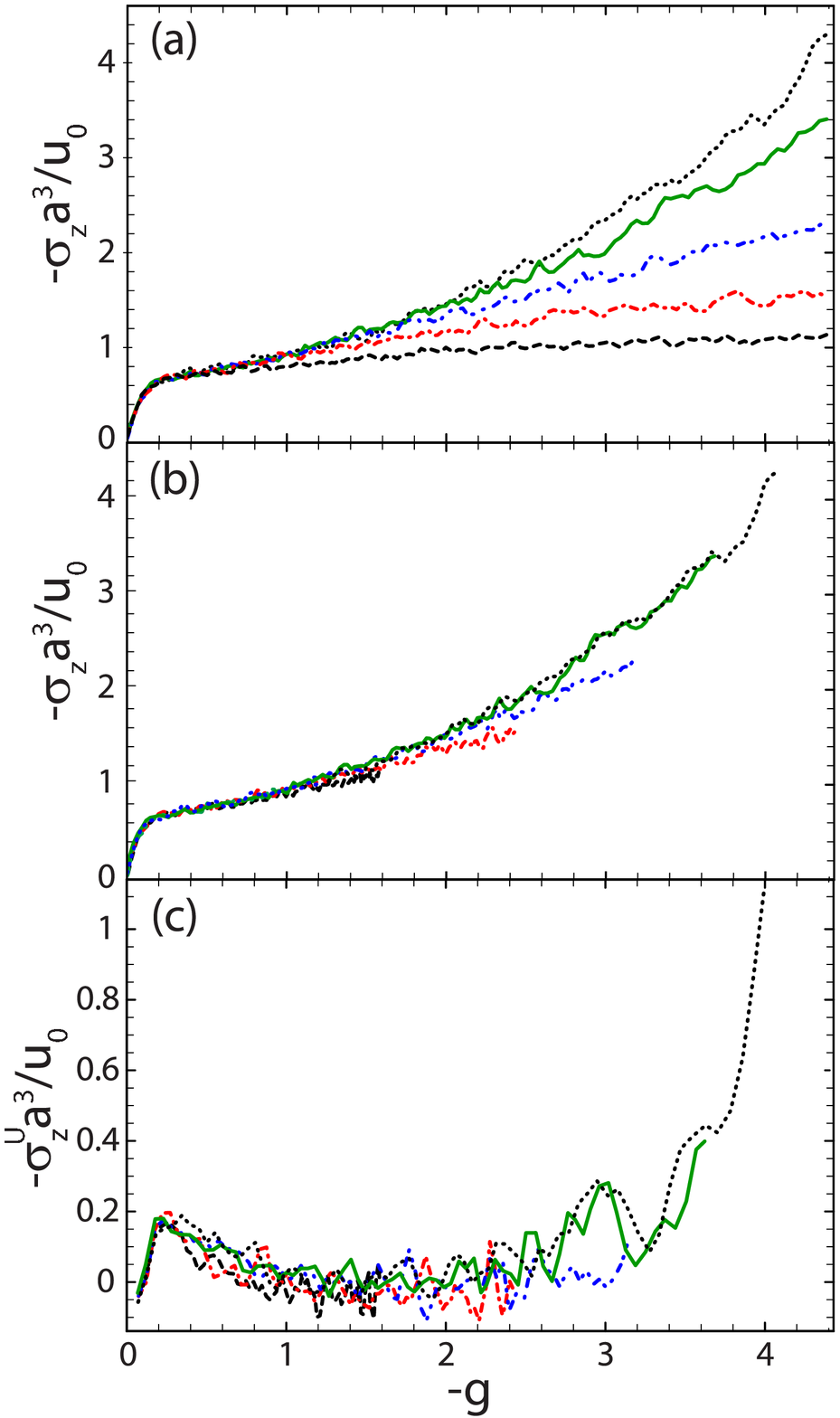}
\caption{
(Color online)
(a) Stress as a function of $g$
during uniaxial compression of $k_{bend}=2.0u_0$ chains
with  $N=350$ (dotted), $88$ (solid), $44$ (dash-dot-dotted),
$22$ (dash-dotted) and $11$ (dashed)
at $T=0.2u_0/k_B$ and $\dot\epsilon=-10^{-5}/\tau_{LJ}$.
(b) Stress replotted against $g_{eff}$.
(c) Energetic stress $\sigma_{z}^{U}$ plotted against $g_{eff}$.
}
\label{fig:Ne22}
\end{figure}

Fig. \ref{fig:collapse2} shows that results for $\sigma_z-\sigma_y$
during plane-strain compression also depend only on microscopic
chain orientation.
When plotted against the macroscopic $|g|$, results for
unentangled chains lie substantially below those for entangled
chains.
When plotted instead against the microscopic orientation function
$g_{eff} = (\lambda_z^c)^2-(\lambda_y^c)^2$,
data for all chains collapse onto a universal curve
(Fig. \ref{fig:collapse2}(b)).
Note that $\lambda_y^c$ decreases by as much as 5\% from $\lambda_y=1$
for the shortest chains, and this affects the data collapse
\cite{foot4}.

The quality of the collapse of the total stress decreases slightly
as the entanglement length decreases.
This is illustrated for $N_e=22$ ($k_{bend}=2.0$) in Fig. \ref{fig:Ne22}.
Results for fully entangled systems ($N \geq 4 N_e $) collapse completely.
Data for smaller $N$ follow the asymptotic curve at small $|g_{eff}|$,
and then drop below it at a $|g_{eff}|$ that decreases with decreasing $N$.
The smallest chains in Fig. \ref{fig:Ne22} and \ref{fig:collapse}(b)
are only a few persistence lengths and may not behave like Gaussian
chains \cite{hoy07}.
However such effects are not large enough to explain
why results for short chains fall below the asymptotic curve in
Fig. \ref{fig:Ne22}(b).

These discrepancies are instead explained by examining the variation with $N$ of the energetic contribution to the stress.
Figure \ref{fig:Ne22}(c) shows $\sigma_{z}^{U}$ plotted against
$g(\lambda_{eff})$.
The initial peak at low $|g_{eff}|$ is nearly independent of
$N$, but the behavior at large $|g_{eff}|$ is not.
There is a sharp rise in $\sigma_z^U$ for fully entangled chains,
that does not occur for $N \leq 44$.
The magnitude of this rise is comparable to the deviation between
results for $N=44$ and the asymptotic curve for entangled chains
in Fig. \ref{fig:Ne22}(a).
These results suggest that while the thermal contribution to
the stress depends only on the chain orientation, the energetic
contribution at large $|g|$ only occurs for entangled chains.
Without the entanglement network, chains can contract along
their tube to eliminate the large energetic stresses.

We have confirmed that increasing the strain rate from
$|\dot\epsilon| = 10^{-5}/\tau_{LJ}$ to $10^{-3}/\tau_{LJ}$
does not change the relation between chain orientation and
stress described in this section.
The main effect is to increase $\lambda_{eff}$ towards $\lambda$
with the increase being more pronounced for shorter chains.
At even higher strain rates, there is almost no relaxation, and
$\lambda_{eff}$ is close to $\lambda$ for all $N$.
This is the regime observed in recent simulations
\cite{lyulin04,lyulin05} with atomistic
potentials, whose greater complexity requires higher strain rates.\newline\newline

\subsection{Dissipative Stresses and Plasticity}
\label{subsec:dissiplastic}

The large value of the hardening modulus and the multiplicative
relation between it and flow stress, suggest that strain hardening
is related to dissipation by plastic rearrangements rather than
stored entropy.
To quantify the rate of plastic deformation $R_P$,
we examine changes in the Lennard-Jones bonds between monomers
over small strain intervals $\delta\epsilon =0.005$.
At the start of the interval, all bonds shorter than $r_c=1.5\sigma$
are identified.
Then the fraction $\delta f$ of these bonds whose length changes by more than
20\% during the interval is evaluated.
This threshold is large enough to exclude changes due to elastic
deformations.
Tests also show that the value of $\delta\epsilon=0.005$ is small
enough that a given atom is unlikely to undergo multiple,
independent bond rearrangements in any interval.
To eliminate activated rearrangements associated with equilibrium
aging, the rate of plasticity during deformation was monitored at $T=0$.

\begin{figure}[htbp]
\includegraphics[width=3.25in]{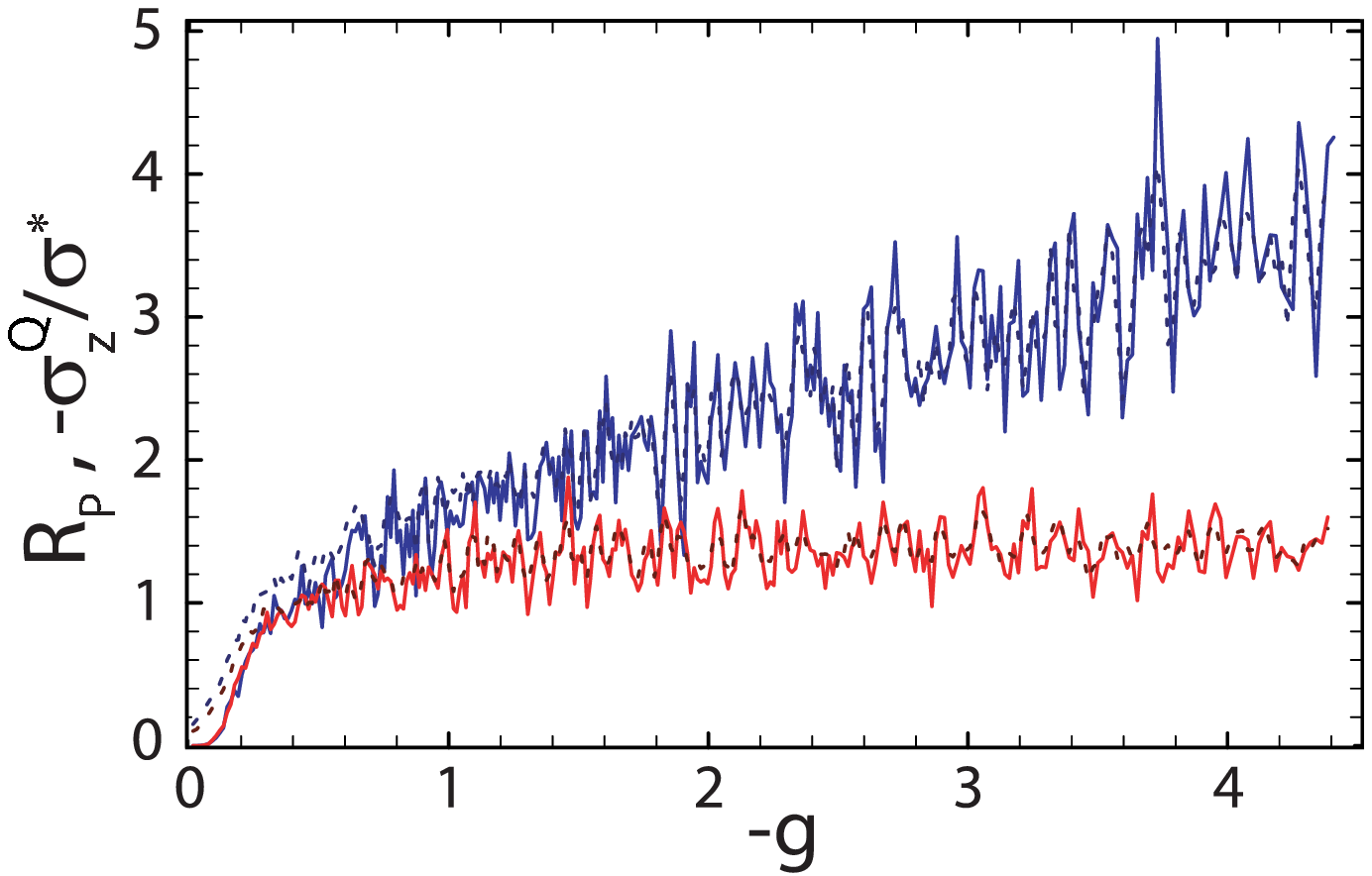}
\caption{
(Color online) Rate of plastic deformation $R_P$ (solid lines)
and normalized dissipative stress (dashed lines)
for uniaxial compression of $k_{bend} = 0.75 u_{0}$ chains
with $N = 350$ (upper curves)
and $N=4$ (lower curves).
Here $T = 0$, $\dot\epsilon = -10^{-4}/\tau_{LJ}$ and
$\sigma^*=1.02u_0/a^3$ for $N=350$ and $1.1u_0/a^3$ for $N=4$.
}
\label{fig:dissdam1}
\end{figure}

Figure \ref{fig:dissdam1}(a) shows the rate of plasticity
$R_P \equiv \delta f/\delta \epsilon$ as a function of $|g|$
during uniaxial compression of fully entangled ($N=350$) and short 
($N=4$) chains with $k_{bend}=0.75$.
The two chain lengths lead to very different curves,
but for both cases $R_P$ is directly proportional to the dissipative
component of the stress.
To illustrate this we also plot $\sigma^Q_z/\sigma^*$ where
$\sigma^{*}$ is the constant of proportionality relating $R_P$ and $\sigma^Q_z$.
Note that even the rapid fluctuations with $|g|$ in the two quantities
are correlated.
These fluctuations are greatly reduced in the total stress,
which can not be made to correlate as well with $R_P$.
The $N=4$ chains exhibit nearly perfect-plastic behavior for $|g| > 1$,
showing that the correlation is not directly related to strain hardening.
The fact that $\sigma^*$ is nearly the same for short chains
that flow at a constant $\tau_{flow}$ and entangled chains that show
significant strain hardening at larger strains is
clear evidence for the close connection between the flow stress
and strain hardening.

\begin{figure}[htbp]
\includegraphics[width=3.25in]{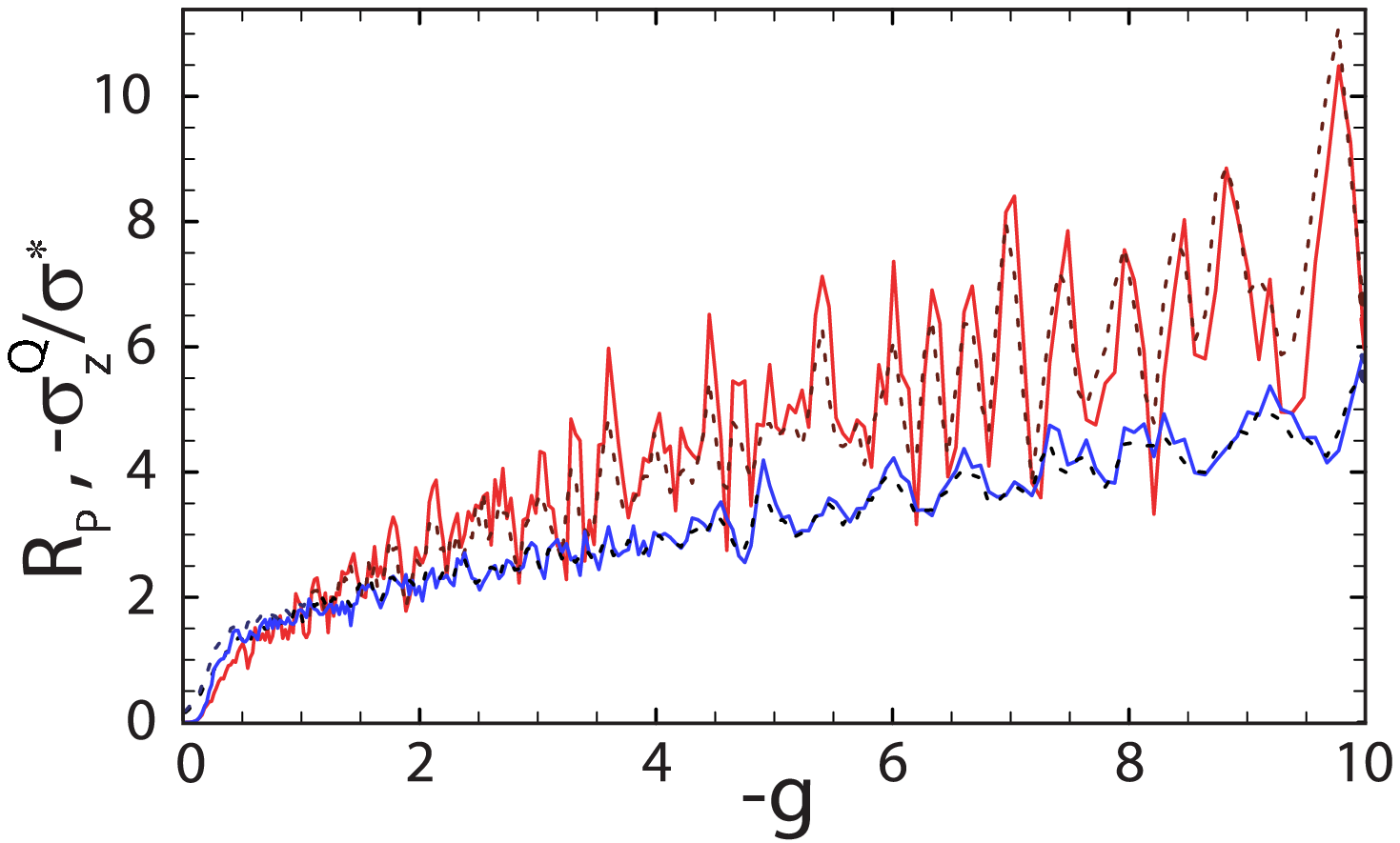}
\caption{
(Color online) Rate of plasticity (solid lines) and dissipative stress
(dashed lines) for plane strain compression of flexible ($k_{bend}=0$)
$N = 500$ chains (lower curves) and semiflexible ($k_{bend}=1.5u_0$)
$N=350$ chains (upper curves).
Here $T = 0$, $\dot\epsilon = -10^{-4}/\tau_{LJ}$ and
$\sigma^*=1.05u_0/a^3$ for $k_{bend}=1.5u_0$ and $0.98u_0/a^3$ for
$k_{bend}=0u_0$.
}
\label{fig:dissdam2}
\end{figure}

In Ref. \cite{hoy07} we showed that $R_P$ and $\sigma_{z}$ were also
correlated for uniaxial compression of chains with $k_{bend} =0$ and
$1.5 u_0$.
Figure \ref{fig:dissdam2} shows that this connection
extends to plane-strain compression.
In all cases studied, $R_P$ tracks both the mean $\sigma_z^Q$ and
local fluctuations.
Moreover, the normalization constants have nearly the same
value within our numerical uncertainties.
Best fits for all $N$, $k_{bend}$ and strain states
range between 0.98 and 1.1$u_0/a^3$.
Since $R_P$ is the rate of rearrangements per LJ bond, $\sigma^*$
should correspond to the density of LJ bonds $\rho_{LJ}$ times
the energy dissipated per bond.
Each atom has on average about 13 LJ neighbors and
each bond is shared by two atoms, so $\rho_{LJ} \sim 6.5 \rho$.
Thus the energy dissipated per bond $\sigma^*/\rho_{LJ}$ is
about a quarter of the binding energy (0.68$u_0$).
Note that this value would change slightly with the threshold
used to define a bond rearrangement and other factors
in the definition of $R_P$, but the result that $\sigma^*$
is similar for all systems with the same $r_c$ is more robust.
Increasing $r_c$ increases the binding energy and also $\sigma^*$.

The same $\sigma^*$ are obtained when the strain rate is
reduced to $10^{-5}/\tau_{LJ}$, but the magnitude of
the fluctuations increases slightly.
Different behavior appears when the strain rate
is increased to $10^{-3}/\tau_{LJ}$. 
Fluctuations are much smaller since there is insufficient
time for stress equilibration.
There is also a decrease in $R_P$, while $\sigma_z^Q$ increases.
This implies that there are fewer plastic rearrangements
involving larger dissipation, presumably because the system
does not have time to minimize the energy.

\section{Summary and Conclusions}
\label{sec:discussconclude}

Extensive simulations of strain hardening were performed for polymer glasses
with a wide range of entanglement densities, chain lengths and temperatures.
As in experiments, we find that the calculated stress-strain curves
of entangled chains
can be fit to expressions derived from entropic network models
(Eq. \ref{eq:origeightchain}).
These models normally treat the flow stress as an independent
parameter that is determined by entirely separate mechanisms.
However, our simulations \cite{hoy06} and experiments \cite{dupaix05} show
that $\tau_{flow}$ and $G_R$ are correlated, and that both
drop linearly to zero as $T$ rises to $T_g$.
This suggests that $\tau_{flow}$ enters multiplicatively
rather than additively, and motivated a simple modification
of the eight-chain model that describes the full temperature
dependence of stress-strain curves.
While this model may prove useful for extrapolating experimental data
at one temperature to all others,
the fit parameters do not appear to have physical significance.
For example, values of $N_e$ are different for uniaxial and plane strain
deformation.
Another difficulty is that the model implies a relation between
the two nonzero stress components in plane strain compression
that is not satisfied by the data.
We are not aware of experimental studies of the transverse stress
$\sigma_y$, but it would be interesting to see if the same
inconsistency could be observed experimentally.

Separate study of the energetic and thermal components of the stress
provided insight into the failures of network models.
The thermal component of the stress $\sigma^Q$ scales nearly
linearly with $|g|$ for all systems.
Even when $h$ is as large as 0.5, the Langevin contribution to strain
hardening (Eq. \ref{eq:origeightchain}) is not evident in $\sigma^Q$.
Instead the rapid rise in $\sigma$ at large $|g|$ is associated
with an increase in the energetic component of stress.
This rise is the dominant factor in fits of $N_e$ to the eight-chain
model.
Since the energetic contributions scale in different ways for uniaxial
and plane strain, fit values of $N_e$ are different for the two strain
states.
Existing experiments have not examined energetic and thermal components
of the stress separately in the strain hardening regime
under isothermal conditions,
but in principle deformation calorimetry experiments could do so.
Our results suggest that a study of trends with entanglement density
would be particularly useful.

Analysis of chain conformations reveals the origin of the energetic
stress.
As strain increases, chains are pulled taut over longer segments.
When the length of straight segments reaches $N_e$, entanglements
limit further straightening.
Additional strain leads to a rapid increase in the tension in covalent bonds
and the energetic component of stress.
Straightening on a scale of order $N_e$ corresponds to large $h$.
Thus even though different strain states lead to different fit
values of $N_e$ in the eight-chain model, both fit values tend to
follow trends in the true $N_e$.
Modern microscopic theories of rubber elasticity \cite{oyerokun04}
incorporate intra- and inter-chain energetic effects due to chain stretching
and orientation, respectively.
Analytic studies based on this approach \cite{chen07b}
may be able to capture the changes in stress and chain
conformation observed in our simulations.

Our simulations reproduce the shape recovery observed in
experiments when strongly deformed,
well-entangled glasses are unloaded and heated slightly above $T_g$
\cite{haward97}.
This relaxation is often invoked as evidence for the entropic stresses
predicted by network models.
As expected from this picture we find a strong correlation between
relaxation of strain and the decay of strain-induced orientational order.
However, we show that the stress associated with shape recovery is
only of order
$\rho_e k_B T$ and thus much too small to account for strain hardening.
This stress was determined by measuring the shear stress in
deformed samples after rapid heating,
and by identifying the shear stress needed to strain
an undeformed sample at the rate observed in shape recovery.
The latter method could also be applied in experiments.

Limited orientation and shape recovery were observed for unentangled
chains even though
entropic models assume that there is no network to impose chain
orientation in such systems.
Significant strain hardening was also found for these unentangled chains.
The stress and orientation follow results for highly entangled
chains at small $|g|$ and saturate at large $|g|$.
The onset of saturation moves to larger $|g|$ as $N$ increases.
As suggested by recent theoretical work \cite{ken} and observed
in our recent simulations \cite{hoy07},
the stress is directly related to effective stretches describing
the microscopic chain orientation $\lambda^c_i$
rather than the macroscopic stretches $\lambda_i$.
Plots of stress against $g(\bar \lambda^c)$ collapse data
for unentangled and highly entangled chains onto a single curve.
Small deviations from this collapse are observed when the energetic
contribution to the stress is large, i.e. when $N_e$ is small
and strain is large.

For both entangled and unentangled chains the thermal contribution to
the stress is directly proportional to the rate of bond rearrangements.
Both the gradual trends and rapid fluctuations in the two quantities
track each other.
The proportionality constant is nearly independent of $N_e$ and
chain length.
The latter result helps to explain the connection between $\tau_{flow}$
and $G_R$ since
short chains shear at a constant stress near $\tau_{flow}$.
There has been great interest recently in plasticity in model atomic
glasses \cite{maloney06,tanguy06,bailey07}.
It would be interesting to check whether the direct correlation between
dissipative stress and bond breaking/reformation holds in such systems \cite{foot6}.

One of the intriguing questions raised by our results is why the
thermal contribution to the stress always rises nearly linearly with $|g|$.
The stress has the functional form expected for the entropy of
Gaussian chains even in the limit $T \rightarrow 0$ where entropic
contributions to the stress must vanish.
One possibility is that entropy enters indirectly.
As $|g|$ increases, the number of conformations available to the chains
is reduced.
The growing constraint on conformations may naturally lead to an increase
in the number of local bond rearrangements that scales with entropy.
This would explain the linear increase in the rate of plasticity
with $|g|$,
and the corresponding increase in $\sigma^Q$.

\section{Acknowledgements}

This material is based upon work
supported by the National Science Foundation under Grant No.\ DMR-0454947.
We thank G. S. Grest for providing equilibrated states and
K. S. Schweizer and D. J. Read for useful discussions.
Simulations were performed with the LAMMPS software package
(http://lammps.sandia.gov).


\begin{thebibliography}{71}
\bibitem[{ken()}]{ken}
\bibinfo{note}{This idea was suggested by K.\ S.\ Schweizer and forms the basis
  of a microscopic theory of strain hardening \cite{chen07b}. The underlying
  physics relates to the theoretical prediction that the bulk modulus of
  strained melts and rubbers increases with chain stretching
  \cite{oyerokun04}.}
\bibitem[{foo({\natexlab{a}})}]{foot1}
\bibinfo{note}{The model is less successful at predicting results for uniaxial
  tension; see e.g. Ref.\ \cite{han99}.}
\bibitem[{foo({\natexlab{b}})}]{foot7}
\bibinfo{note}{Note that the $N_e=71$ chains appear to deform slightly
  superaffinely (by $<$0.5\%) at moderate strains. This is a finite size
  effect. It arises from the use of the formula $R_{aff}(n) =
  \lambda_{chain}R_{0}(n)$, which assumes the averages over chains of the
  Cartesian components of $R_0^2(n)$ are isotropic. In a finite system there
  are always small deviations from anisotropy. Correcting for them does not
  change other features in the plot.}
\bibitem[{foo({\natexlab{c}})}]{foot3}
\bibinfo{note}{Of course $R(n)$ is an average over all points on the chain. One
  might imagine that there are localized entanglement points that deform more
  affinely or that there is a distribution of entanglement separations as seen
  in simulations \cite{tzoumanekas06}. While this might preserve the idea of an
  affinely deformed network, the discrepancy from the eight-chain model would
  remain.}
\bibitem[{foo({\natexlab{d}})}]{foot4}
\bibinfo{note}{Because the magnitude of the stress difference is smaller, the
  fluctuations between curves appear greater than in Fig. \ref{fig:collapse}.}
\end{thebibliography}
\end{document}